\documentclass[%
 reprint,
 superscriptaddress,
 amsmath,amssymb,
 aps,
 prb,
 nofootinbib
]{revtex4-2}
   
\usepackage{graphicx}
\usepackage{siunitx}
\usepackage{hyperref}
\hypersetup{colorlinks=true,citecolor=blue, urlcolor=blue, linkcolor=blue}
\usepackage[pagewise]{lineno}
\begin{document}


\title{Superconducting Diode Effect and Large Magnetochiral Anisotropy\\ in T$_d$-MoTe$_2$ Thin Film}

\author{Wan-Shun Du}
\thanks{These authors contributed equally to this work.}
\affiliation{Shenzhen Institute for Quantum Science and Engineering, Southern University of Science and Technology, Shenzhen 518055, China}
\affiliation{Department of Physics, Southern University of Science and Technology, Shenzhen 518055, China}
\affiliation{Department of Physics, The Hong Kong University of Science and Technology, Clear Water Bay, Kowloon, Hong Kong, China}
\author{Weipeng Chen}
\thanks{These authors contributed equally to this work.}
\affiliation{Shenzhen Institute for Quantum Science and Engineering, Southern University of Science and Technology, Shenzhen 518055, China}
\affiliation{International Quantum Academy, Shenzhen 518048, China}
\affiliation{Guangdong Provincial Key Laboratory of Quantum Science and Engineering, Southern University of Science and Technology, Shenzhen 518055, China}
\author{Yangbo Zhou}
\thanks{These authors contributed equally to this work.}
\affiliation{Department of Materials Science, School of Physics and Materials Science, Nanchang University, Nanchang, Jiangxi, 330031, China}
\affiliation{Jiangxi Key Laboratory for Two-Dimensional Materials, Nanchang University, Nanchang, Jiangxi, 330031, China}
\author{Tengfei Zhou}
\thanks{These authors contributed equally to this work.}
\affiliation{Shenzhen Institute for Quantum Science and Engineering, Southern University of Science and Technology, Shenzhen 518055, China}
\author{Guangjian Liu}
\affiliation{Department of Materials Science, School of Physics and Materials Science, Nanchang University, Nanchang, Jiangxi, 330031, China}
\affiliation{Jiangxi Key Laboratory for Two-Dimensional Materials, Nanchang University, Nanchang, Jiangxi, 330031, China}
\author{Zongteng Zhang}
\affiliation{Department of Physics, Southern University of Science and Technology, Shenzhen 518055, China}
\author{Zichuan Miao}
\affiliation{Department of Materials Science $\&$ Metallurgy, University of Cambridge, 27 Charles Babbage Road, Cambridge, CB3 0FS, United Kingdom}
\author{Hao Jia}
\affiliation{Shenzhen Institute for Quantum Science and Engineering, Southern University of Science and Technology, Shenzhen 518055, China}
\affiliation{International Quantum Academy, Shenzhen 518048, China}
\affiliation{Guangdong Provincial Key Laboratory of Quantum Science and Engineering, Southern University of Science and Technology, Shenzhen 518055, China}
\author{Song Liu}
\affiliation{Shenzhen Institute for Quantum Science and Engineering, Southern University of Science and Technology, Shenzhen 518055, China}
\affiliation{International Quantum Academy, Shenzhen 518048, China}
\affiliation{Guangdong Provincial Key Laboratory of Quantum Science and Engineering, Southern University of Science and Technology, Shenzhen 518055, China}
\author{Yue Zhao}
\affiliation{Shenzhen Institute for Quantum Science and Engineering, Southern University of Science and Technology, Shenzhen 518055, China}
\affiliation{Department of Physics, Southern University of Science and Technology, Shenzhen 518055, China}
\author{Zhensheng Zhang}
\affiliation{Shenzhen Institute for Quantum Science and Engineering, Southern University of Science and Technology, Shenzhen 518055, China}
\affiliation{International Quantum Academy, Shenzhen 518048, China}
\author{Tingyong Chen}
\affiliation{Shenzhen Institute for Quantum Science and Engineering, Southern University of Science and Technology, Shenzhen 518055, China}
\affiliation{International Quantum Academy, Shenzhen 518048, China}
\author{Ning Wang}
\affiliation{Department of Physics, The Hong Kong University of Science and Technology, Clear Water Bay, Kowloon, Hong Kong, China}
\author{Wen Huang}
\email{huangw3@sustech.edu.cn}
\affiliation{Shenzhen Institute for Quantum Science and Engineering, Southern University of Science and Technology, Shenzhen 518055, China}
\affiliation{International Quantum Academy, Shenzhen 518048, China}
\affiliation{Guangdong Provincial Key Laboratory of Quantum Science and Engineering, Southern University of Science and Technology, Shenzhen 518055, China}
\author{Zhen-Bing Tan}
\email{tanzb@sustech.edu.cn}
\affiliation{Shenzhen Institute for Quantum Science and Engineering, Southern University of Science and Technology, Shenzhen 518055, China}
\affiliation{International Quantum Academy, Shenzhen 518048, China}
\author{Jing-Jing Chen}
\email{chenjj3@sustech.edu.cn}
\affiliation{Shenzhen Institute for Quantum Science and Engineering, Southern University of Science and Technology, Shenzhen 518055, China}
\affiliation{International Quantum Academy, Shenzhen 518048, China}
\affiliation{Guangdong Provincial Key Laboratory of Quantum Science and Engineering, Southern University of Science and Technology, Shenzhen 518055, China}
\author{Da-Peng Yu}
\email{yudp@sustech.edu.cn}
\affiliation{Shenzhen Institute for Quantum Science and Engineering, Southern University of Science and Technology, Shenzhen 518055, China}
\affiliation{International Quantum Academy, Shenzhen 518048, China}
\affiliation{Guangdong Provincial Key Laboratory of Quantum Science and Engineering, Southern University of Science and Technology, Shenzhen 518055, China}

\date{\today}

\begin{abstract}
In the absence of time-reversal invariance, metals without inversion symmetry may exhibit nonreciprocal charge transport --- a magnetochiral anisotropy that manifests as unequal electrical resistance for opposite current flow directions. If superconductivity also sets in, the charge transmission may become dissipationless in one direction while remaining dissipative in the opposite, thereby realizing a superconducting diode. Through both direct-current and alternating-current measurements, we study the nonreciprocal effects in thin films of the noncentrosymmetric superconductor T$_d$-MoTe\textsubscript{2} with disorders. We observe nonreciprocal superconducting critical currents with a diode efficiency close to 20\%~, and a large magnetochiral anisotropy coefficient up to $\SI{5.9e8}{\per\tesla\per\ampere}$, under weak out-of-plane magnetic field in the millitesla range.  The great enhancement of rectification efficiency under out-of-plane magnetic field is likely abscribed to the vortex ratchet effect, which naturally appears in the noncentrosymmetric superconductor with disorders. Intriguingly, unlike the finding in Rashba systems, the strongest in-plane nonreciprocal effect does not occur when the field is perpendicular to the current flow direction. We develop a phenomenological theory to demonstrate that this peculiar behavior can be attributed to the asymmetric structure of spin-orbit coupling in T$_d$-MoTe\textsubscript{2}. Our study highlights how the crystallographic symmetry critically impacts the nonreciprocal transport, and would further advance the research for designing the superconducting diode with the best performance.
\end{abstract}

\maketitle
\section{Introduction}
The nonreciprocal effect where charge transport exhibits directional dependence, such as in p-n junction, is fundamental to modern electronics. Recently, the nonreciprocal charge responses in superconducting systems have stimulated a great deal of interest ~\cite{wakatsuki2017nonreciprocal, qinSuperconductivityChiralNanotube2017,yasuda2019nonreciprocal,itahashi2020nonreciprocal,zhang2020nonreciprocal,itahashiQuantumClassicalRatchet2020,ideueGiantNonreciprocalMagnetotransport2020}. In particular, superconducting diodes that manifest distinct critical currents in different directions have been demonstrated~\cite{ando2020observation, baumgartner2022supercurrent,wu2022field,lin2022zero, pal2022Josephson,houUbiquitousSuperconductingDiode2023,bauriedl2022supercurrent,yunMagneticProximityinducedSuperconducting2023}. Such a diode supports dissipationless charge transmission in one direction while exhibiting finite resistance in the opposite, paving the way for future applications in non-dissipative electronics.

The superconducting diode effect (SDE) is a descendant of a larger set of phenomena called magnetochiral
anisotropy (MCA)~\cite{rikken1997observation, rikken2001electrical}. It may occur in bulk crystals when spatial inversion and time reversal symmetries are both broken~\cite{tokura2018nonreciprocal}. A case in point is metals with noncentrosymmetric structure, where time reversal invariance can be broken by the application of an external magnetic field. The nonreciprocity can be described by a sample resistance that depends on the directions of the current flow and magnetic field \cite{rikken2005magnetoelectric},
\begin{equation}\label{eq:1}
    R = R_0(1+\gamma \mu_0H I) \,,
\end{equation}
where $R_0$ is the resistance at zero field, $I$ is the current, $\mu_0H$ represents the magnetic field, and $\gamma$ is a coefficient characterizing the strength of the MCA. While the MCA coefficient is usually small in normal conductors due to the small spin–orbit coupling ($\sim$ meV) relative to the Fermi energy ($\sim$ eV)~\cite{ideue2017bulk, tokura2018nonreciprocal}. It is argued to be markedly amplified by superconducting fluctuations slightly above the superconducting transition because of the replacement of the energy denominator from the Fermi energy to the superconducting gap ($\sim$ meV)~\cite{hoshino2018nonreciprocal}, as has been demonstrated in noncentrosymmetric superconductors MoS$_2$, SrTiO$_3$, NbSe$_2$, as well as in the Bi$_2$Te$_3$/FeTe heterostructure~\cite{wakatsuki2017nonreciprocal, itahashi2020nonreciprocal, yasuda2019nonreciprocal, zhang2020nonreciprocal}. 

Another way to enhance rectification is to engineer the ratchet motion of vortices  in a superconductor with artificial asymmetric pinning potentials~\cite{leeReducingVortexDensity1999,villegasSuperconductingReversibleRectifier2003,vandevondelVortexRectificationEffectsFilms2005,villegasExperimentalRatchetEffect2005,palauGuidedVortexMotion2012,lyu2021superconducting}. Recent theory has revealed that the vortex ratchet effect can appear as a consequence of disorders in noncentrosymmetric superconductors whose asymmetric crystal structure intrinsically produces asymmetric pinning potentials for vortices, which is distinct from the previously discussed artificially developed inversion broken environment~\cite{hoshino2018nonreciprocal}. Such nonreciprocal charge transport has been reported in the trigonal crystals such as two-dimensional MoS\textsubscript{2}~\cite{itahashiQuantumClassicalRatchet2020}, NbSe\textsubscript{2}~\cite{zhang2020nonreciprocal} and bulk PbTaSe\textsubscript{2}~\cite{ideueGiantNonreciprocalMagnetotransport2020}, and in the few-layer T$_d$-MoTe\textsubscript{2}~\cite{wakamura2023gate} with lower symmetry than trigonal symmetry. In most of these studies, the nonreciprocal charge transport from the vortex ratchet effect is detected by the second-harmonic AC measurement, but the nonreciprocal supercurrent is not realized due to the small rectification efficiency. As the disorders play important roles as pinning centers for vortices, it is particularly called for to investigate whether it is possible to enhance the rectification efficiency by controlling the disorders in such noncentrosymmetric superconductors, thereby achieving the superconducting diode. 

In this work, we study the nonreciprocal transport in thin films of the noncentrosymmetric superconductor T$_d$-MoTe\textsubscript{2} with disorders induced by argon etching. This compound is a type-II Weyl semimetal with a superconducting transition temperature $T_c \approx$ 0.1 K \cite{sun2015prediction,qi2016superconductivity,zhang2016raman}. Our direct-current (DC) measurements provide clear evidence for critical current nonreciprocity in the presence of weak out-of-plane magnetic field, with a diode efficiency $2\times(I_{c,+} - |I_{c,-}|)/(I_{c,+} + |I_{c,-}|)$ around 20\%. The alternating-current (AC) measurements unveil strong MCA with $\gamma$ reaching up to \SI{5.9e8}{\per\tesla\per\ampere} --- much larger than previous reports in other systems. The great enhancement of rectification efficiency under out-of-plane magnetic field is likely attributed to the vortex ratchet effect. The large MCA also appears under in-plane magnetic fields, but unlike in some Rashba superconductors~\cite{wakatsuki2018nonreciprocal, hoshino2018nonreciprocal,yasuda2019nonreciprocal,itahashi2020nonreciprocal,ando2020observation}, the strongest MCA does not occur when the field is perpendicular to the current flow direction. On the basis of a phenomenological theory, we show that this intriguing behavior has its origin in the peculiar form of the spin-orbit coupling in T$_d$-MoTe\textsubscript{2}.

\section{Results}
\subsection{Transport properties}
\begin{figure}
\includegraphics{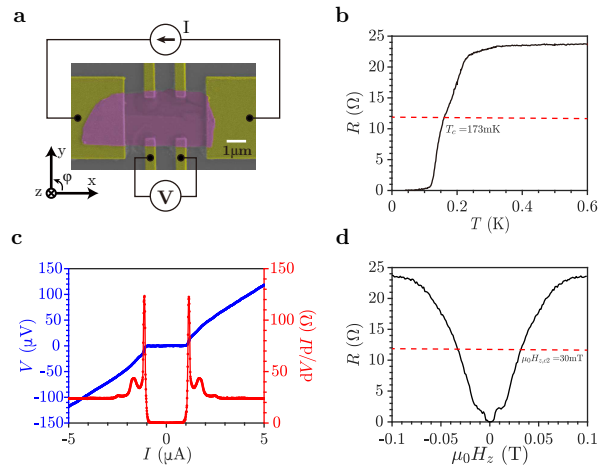}
\caption{\textbf{T$_d$-$\text{MoTe}_2$ device and superconductivity.} (a) False-color SEM image of the $\text{MoTe}_2$ device and the measurement setup with bias-current and monitored voltage. x, y, z and $\mathbf{\varphi}$ indicate the coordinate system defined in the measurement. (b) Temperature-dependent resistance from \SIrange{600}{10}{\milli\kelvin}. The red dashed line denotes the resistance reaching half of that in normal state. (c) Voltage-current ($V -I$) characteristic and the corresponding differential resistance $\mathrm{d}V/\mathrm{d}I$ at \SI{10}{\milli\kelvin}. (d) Magnetoresistance under out-of-plane field at \SI{10}{\milli\kelvin}.  \label{fig:1}}
\end{figure}

Fig.~\ref{fig:1}a shows the scanning electron microscope (SEM) image of the measured T$_d$-MoTe\textsubscript{2} device with an area of \SI{13.17}{\square\micro\meter}. Thin microflakes were obtained by mechanical exfoliation from bulk single crystals and then transferred onto a Si substrate capped with a \SI{285}{\nano\meter} thick $\text{SiO}_2$ layer. To introduce defects as vortex pins, the sample was pre-etched by argon (Ar) plasma for \SI{400}{\second} and was overall thinned from about \SI{62}{\nano\meter} to \SI{43}{\nano\meter} (Supplemental Material Fig. S1 (SM Fig. S1)~\cite{SeeSupplementalMaterial}). Electrical contacts were then defined using electron beam lithography. To realize ohmic contact between the $\text{MoTe}_2$ sample and electrodes (5/150\SI{}{\nano\meter} Ti/Au), an \textit{in situ} Ar plasma etch for \SI{80}{\second} was employed prior to metal deposition. Standard four-terminal measurements were performed in a dilution refrigerator with a base temperature of \SI{10}{\milli\kelvin}. 

The Hall measurement in normal state shows dominant electron-type transport, with the carrier density calculated to be \SI{7e20}{\per\cubic\centi\meter} and the mobility $\mu=$ \SI{31}{\square\centi\meter\per\volt\per\second}  at \SI{10}{\milli\kelvin} (SM Fig. S2~\cite{SeeSupplementalMaterial}). These characteristics are comparable to what was reported in CVD samples~\cite{cui2019transport}. The carrier density is approximately one order of magnitude larger than that in bulk samples~\cite{qi2016superconductivity,zhouHallEffectColossal2016}, which is probably due to the increased electron doping  by Te vacancies created by argon etch~\cite{choTeVacancydrivenSuperconductivity2017}. Likewise, the mobility is much lower than the reported value~\cite{zhouHallEffectColossal2016}. Both the higher carrier density and lower mobility suggest a considerable defect concentration induced by argon plasma etching, although the plasma can etch the 2D materials layer-by-layer in some condition~\cite{liu2013layer}.

Fig.~\ref{fig:1}b shows the temperature dependence of the sheet resistance $R$, measured under zero-field. The experiment was performed using standard lock-in techniques with a small AC current excitation $I_{ac} = \SI{40}{\nano\ampere}$ at a frequency of $\SI{73}{\hertz}$. As the sample is cooled down, the resistance suddenly drops at $\SI{0.23}{\kelvin}$, indicating the onset of superconductivity; zero-resistance is achieved when the temperature is lower than $T_{c,0} = \SI{0.09}{\kelvin}$. Empirically, the superconducting phase transition temperature $T_{c,r}$ is determined to be \SI{0.173}{\kelvin}, selected by a reduced resistance $r = R / R_{0.6\text{K}} = 0.5$.  Note that the resistance undergoes multiple steep falls before dropping to zero, which could be attributed to the inhomogeneity in the sample due to argon etching~\cite{fatemiElectricallyTunableLowdensity2018}. Fig.~\ref{fig:1}c shows the voltage-current ($V$ - $I$) characteristics and the corresponding differential resistance ($\mathrm{d}V/\mathrm{d}I$) at \SI{10}{\milli\kelvin}. Zero-voltage is observed when the injected DC current satisfies $|I|<\SI{0.9}{\micro\ampere}$, signifying a superconducting state. At larger currents ($|I|>\SI{2.5}{\micro\ampere}$), the sample returns to normal state and the $V$ - $I$ curve follows a straight line with a slope of \SI{23.8}{\ohm}. In addition, we can see multiple peaks in the $\mathrm{d}V/\mathrm{d}I$ - $I$ curve in association with the previously mentioned sample inhomogeneity. 

We further characterize the suppression of the superconductivity by applying an out-of-plane magnetic field $\mu_0H_z$. As can be seen in Fig.~\ref{fig:1}d, the sample resistance gradually develops as the field increases above \SI{3}{\milli\tesla} and tends to saturate until around \SI{0.1}{\tesla}.  The out-of-plane upper critical field $\mu_0H_{c2,z}$ is determined to be \SI{30}{\milli\tesla}, at which half of the normal state resistance is recovered. Such a significantly broadened superconducting transition is attributed to the vortex effect, which is frequently observed in high $T_c$ cuprates. In addition, we have also confirmed the 2D nature of the superconductivity by checking the temperature dependences of the out-of-plane and in-plane upper critical fields, and the 2D Berezinskii–Kosterlitz–Thouless (BKT) transition (SM Fig. S3~\cite{SeeSupplementalMaterial}).
\begin{figure*}
\includegraphics{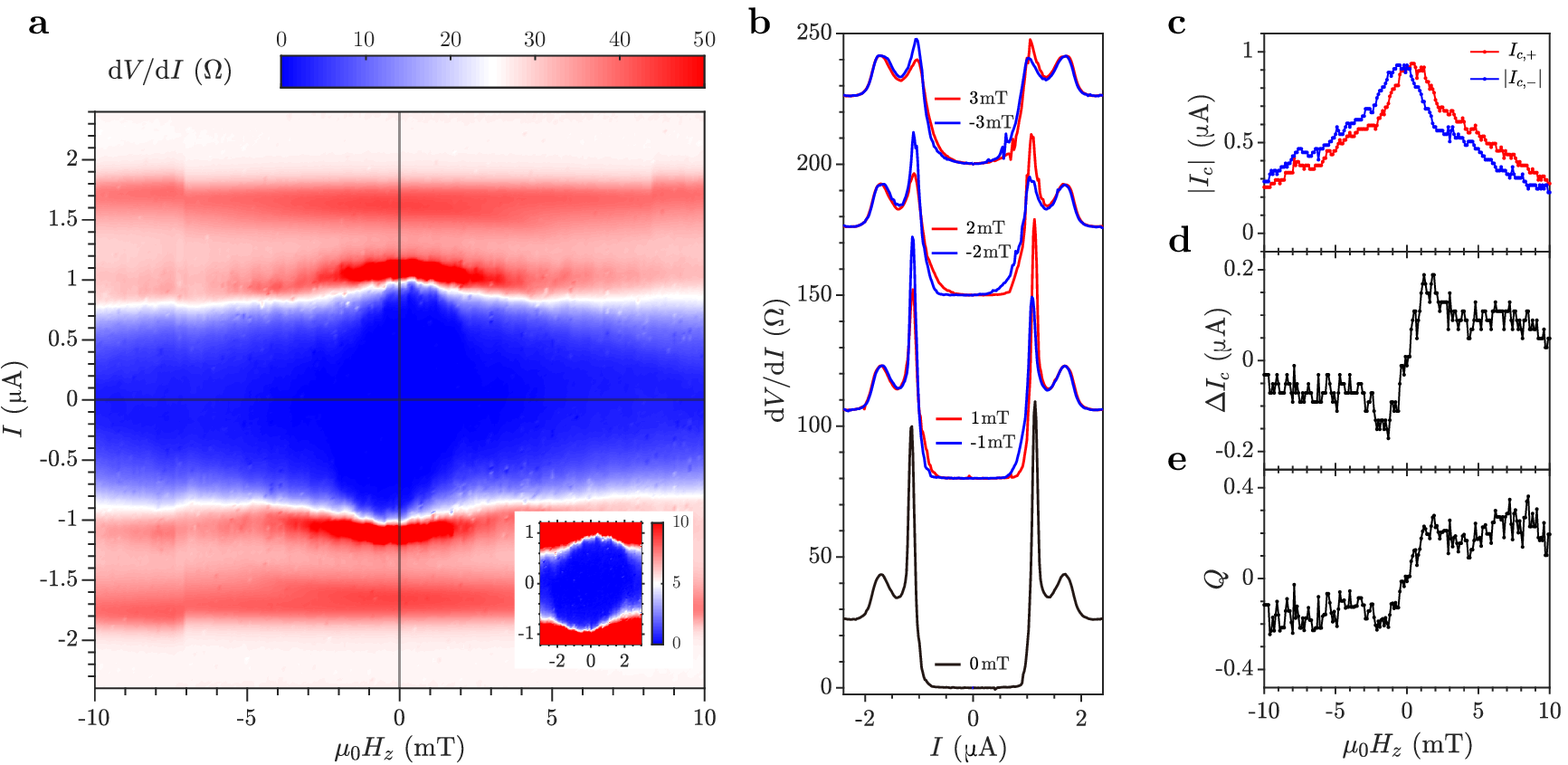}
\caption{\textbf{Nonreciprocal superconducting transport under out-of-plane magnetic field.} (a) Differential resistance $\mathrm{d}V/\mathrm{d}I$ as a function of out-of-plane magnetic field $\mu_0 H_z$ and DC current $I$. (inset) Zoom-in around $|\mu_0H_z|<\SI{2.5}{\milli\tesla}$ and $|I|<\SI{1.25}{\micro\ampere}$.  (b) Linecuts of $\mathrm{d}V/\mathrm{d}I$ at $\mu_0 H_z = $ \SIlist[list-units=single,list-final-separator={, }]{0;\pm 1; \pm 2; \pm 3}{\milli\tesla}, extracted from (a). The curves are slightly shifted upward for clarification. (c) Nonreciprocal critical current $|I_c|$, (d) difference of the critical current $\Delta I_c=I_{c,+} - |I_{c,-}|$, and (e) supercurrent rectification efficiency $Q=\frac{\Delta I_c}{(I_{c,+} + |I_{c,-}|)/2}$, plotted as a function of magnetic field.}\label{fig:2}
\end{figure*}

\subsection{Nonreciprocal superconducting transport}
Next, we focus on the nonreciprocal superconducting transport under out-of-plane magnetic field. Fig.~\ref{fig:2}a shows the differential resistance colormap as a function of DC current $I$ and out-of-plane magnetic field $\mu_0 H_z$, exhibiting clear asymmetry with respect to both $I$ and $\mu_0H_z$. With the zero-field $\mathrm{d}V/\mathrm{d}I$ plotted as a reference, Fig.~\ref{fig:2}b shows the $\mathrm{d}V/\mathrm{d}I$ linecuts for $\mu_0 H_z$ =  \SIlist[list-units=single,list-final-separator={, }]{\pm 1; \pm 2; \pm 3}{\milli\tesla}, revealing a clear nonreciprocal magneto-transport. The nonreciprocity also manifests in the unequal superconducting critical currents in opposite directions, $I_{c,+}$ and $|I_{c,-}|$, as shown in Fig.~\ref{fig:2}c. The relative magnitude between $I_{c,+}$ and $| I_{c,-} |$ reverses when the external field reverses direction. This is known as the superconducting diode effect, in which an injected current in the range between $I_{c,+}$ and $| I_{c,-} |$ can only flow in one direction without dissipation. Fig.~\ref{fig:2}d shows the nonreciprocal component of the critical current $\Delta I_c = I_{c,+} - |I_{c,-}|$ as a function of $\mu_0 H_z$. $\Delta I_c$ reaches its maximal value of $\SI{0.2}{\micro\ampere}$ at about $\SI{2}{\milli\tesla}$, and then decreases to about zero as the magnetic field increases further to $\SI{10}{\milli\tesla}$. The quality of the diode effect is usually described in terms of the supercurrent rectification efficiency \cite{he2022phenomenological}, namely, $Q = \frac{\Delta I_{c}}{(I_{c,+} + |I_{c,-}|)/2}$, which is presented in Fig.~\ref{fig:2}e. The rectification efficiency first increases with field, and then reaches a stable value of about $20\%$ for $\SI{2}{\milli\tesla}<\mu_0H_z<\SI{10}{\milli\tesla}$. Such a large value is much higher than those ($5\%$) observed in Refs.~\cite{ando2020observation,wu2022field}, and is comparable to those in some devices reported recently ~\cite{bauriedl2022supercurrent,yunMagneticProximityinducedSuperconducting2023,baumgartner2022supercurrent}.

Similar measurements on another two intrinsic (without argon etch) MoTe$_2$ devices as control experiments were also performed. As shown in SM Fig. S4~\cite{SeeSupplementalMaterial}, no obvious superconducting diode effect can be observed in the intrinsic samples, indicating the important role of disorders in the system. With a magnetic field applied, vortices  penetrate the superconductor and are usually trapped by pinning potentials induced by defects. Recent theory has revealed that vortices driven by the external current feel an asymmetric pinning potential in the superconductor without inversion center, acquiring unequal velocity depending on the current direction~\cite{hoshino2018nonreciprocal}. It is noted that this vortex ratchet effect naturally appears as a consequence of disorders in noncentrosymmetric system, and the consequent nonreciprocal charge transport has been reported in the trigonal crystals such as two-dimensional MoS\textsubscript{2}~\cite{itahashiQuantumClassicalRatchet2020}, NbSe\textsubscript{2}~\cite{zhang2020nonreciprocal} and bulk PbTaSe\textsubscript{2}~\cite{ideueGiantNonreciprocalMagnetotransport2020}, and in the few-layer T$_d$-MoTe\textsubscript{2}~\cite{wakamura2023gate} with lower symmetry than trigonal symmetry. Here, the rectification efficiency is greatly enhanced by introducing more vortex pins, thereby realizing the superconducting diode effect.

\begin{figure*}
\includegraphics{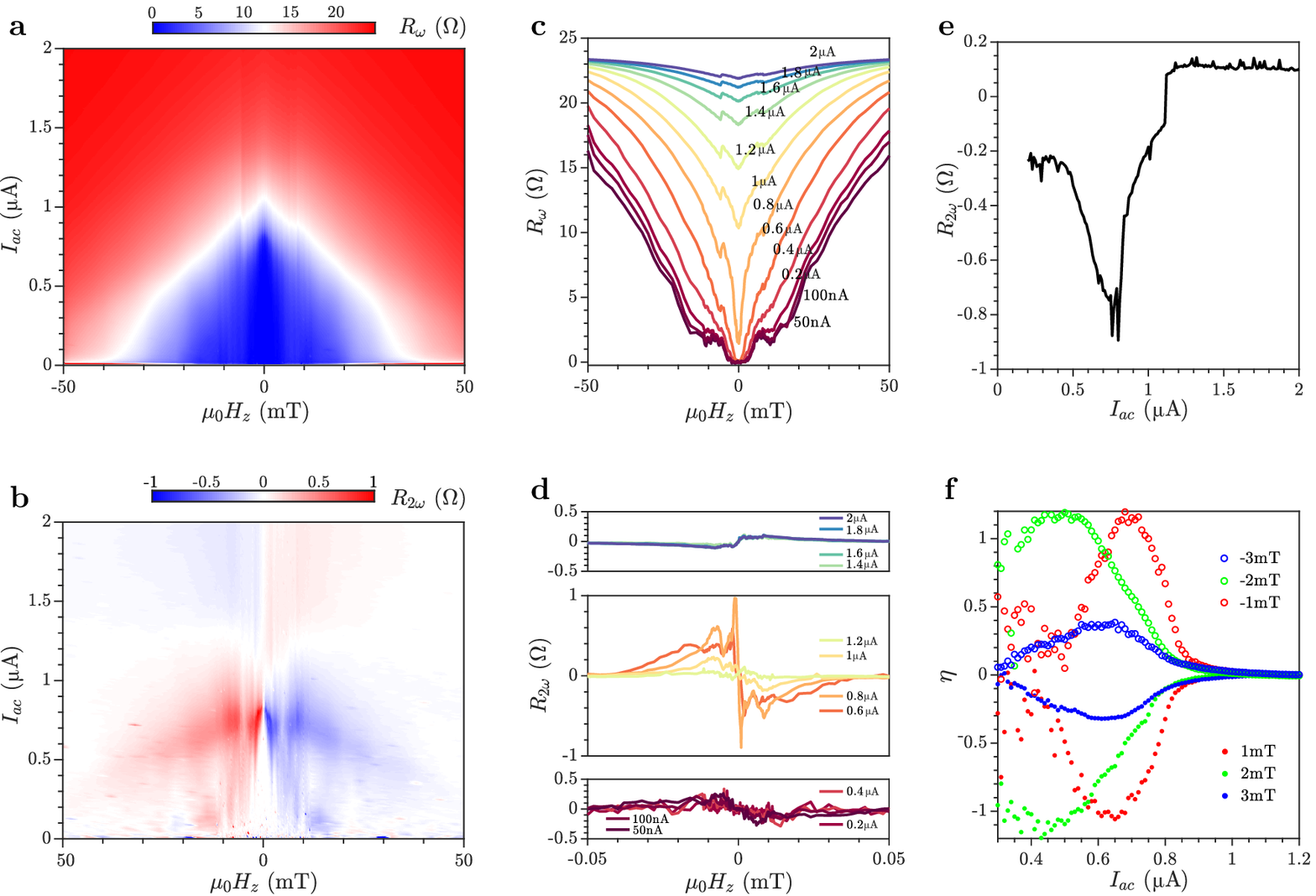}
\caption{\textbf{Out-of-plane magnetochiral anisotropy.} (a) First harmonic resistance $R_\omega$ and (b) second harmonic resistance $R_{2\omega}$ as a function of $\mu_0 H_z$ and AC current $I_{ac}$. Linecuts of (c) $R_{\omega}$ and (d) $R_{2\omega}$ as a function of magnetic field at various AC currents. (e) The maximal $R_{2\omega}$ signal under positive magnetic field plotted as a function of AC current. (f) {$\eta$, the ratio of the nonlinear resistance $2\times R_{2\omega}$ to the linear resistance $R_{\omega}$,}  as a function of $I_{ac}$ at $\mu_0 H_z = $ \SIlist[list-units=single, list-final-separator={, }]{\pm 1; \pm 2; \pm 3}{\milli\tesla}. }\label{fig:3}
\end{figure*}

\subsection{Out-of-plane magnetochiral anisotropy}
To further investigate the properties of the nonreciprocal transport, we performed the second-harmonic AC measurement. The measurement on the harmonic resistances was performed by sweeping the AC current $I_{ac}$ without DC current offset under various out-of-plane magnetic field. Figs.~\ref{fig:3}a and \ref{fig:3}b show the colormaps of the first and second harmonic resistances, $R_\omega$ and $R_{2\omega}$, as a function of $\mu_0 H_z$ and $I_{ac}$ at \SI{10}{\milli\kelvin}, respectively. Here, $R_\omega$ corresponds to linear resistance $R_0$ and $R_{2\omega}$ represents the half of the nonlinear resistance which is proportional to the current and the magnetic field ($R_{2\omega} = \frac{R_0}{2}\gamma\mu_0HI_p$, $I_p$ is the amplitude of the AC current)~\cite{itahashi2020nonreciprocal,ando2020observation}. It can be seen that $R_\omega$ is symmetric with respect to the magnetic field while $R_{2\omega}$ is antisymmetric, which is characteristic of the MCA~\cite{tokura2018nonreciprocal}. It should be noted that the nonlinear current-voltage curve of a superconductor may also produce the second harmonic signals. If so, the $R_{2\omega}$ should be symmetric in the magnetic filed, and exists even in the absence of the magnetic field, inconsistent with the observations.

Figs.~\ref{fig:3}c and \ref{fig:3}d show field dependences of $R_{\omega}$ and $R_{2\omega}$ at various AC currents extracted from Figs.~\ref{fig:3}a and \ref{fig:3}b, respectively.  It is clear that the $R_{2\omega}$ appears in the superconducting transition regime. To see the influence of the current, the maximal $R_{2\omega}$ signal under positive magnetic field is plotted as a function of the AC current in Fig.~\ref{fig:3}e, showing a peak structure, which is similar to that of the vortex ratchet effect observed in previous studies~\cite{villegasSuperconductingReversibleRectifier2003,villegasExperimentalRatchetEffect2005,itahashiQuantumClassicalRatchet2020,zhang2020nonreciprocal}. As one can see, the $R_{2\omega}$ signals appear as the AC current is larger than about \SI{0.2}{\micro\ampere}, but don't change until \SI{0.5}{\micro\ampere}. In this low current regime, the trapped vortices cannot be depinned at all, causing a suppressed nonreciprocal response. This weak nonreciprocal transport may come from interstitial vortices moving away from the pin centers~\cite{zhuControllableStepMotors2003}. As the AC current further increases, vortices are driven and acquire an increasing net velocity, giving rise to rapid increase of the $R_{2\omega}$~\cite{zhuControllableStepMotors2003,villegasSuperconductingReversibleRectifier2003,villegasExperimentalRatchetEffect2005}. When the AC current is larger than \SI{0.8}{\micro\ampere}, the $R_{2\omega}$ decreases due to the effective weakening of the asymmetric pinning potential~\cite{zhuControllableStepMotors2003,villegasSuperconductingReversibleRectifier2003,villegasExperimentalRatchetEffect2005,itahashiQuantumClassicalRatchet2020,zhang2020nonreciprocal}. Note that the $R_{\omega}$ is larger than half of the normal state resistance when $I_{ac}>\SI{1}{\micro\ampere}$, so the quenching of the superconductivity could also be considered. When the AC current is larger than \SI{1.1}{\micro\ampere}, a sign reversal of the $R_{2\omega}$ signals is observed, implying a new mechanism of the nonrecipracal transport. In such a high resistance region, the amplitude fluctuation of the superconducting order parameter cannot be neglected, so the paraconductivity may dominate the nonreciprocal transport~\cite{ideueGiantNonreciprocalMagnetotransport2020,wakatsuki2017nonreciprocal,hoshino2018nonreciprocal}. However, the paraconductivity induced nonreciprocity is much smaller than the vortex induced signal in the present system. The crossover behavior from the paraconductivity origin to the vortex origin is also observed by warming above the mean-field transition temperature where the amplitude fluctuation dominates the nonreciprocal transport (SM Fig. S5~\cite{SeeSupplementalMaterial}), consistent with the results in previous studies\cite{itahashi2020nonreciprocal,wakatsuki2017nonreciprocal,hoshino2018nonreciprocal}.

Another notable feature of the second harmonic resistance is the oscillating behavior with the magnetic field, as seen in Fig.~\ref{fig:3}d ($I_{ac}=\SI{0.8}{\micro\ampere}$ and $\SI{1}{\micro\ampere}$). Such an oscillating behavior probably comes from the vortex matching effect\cite{itahashi2020nonreciprocal,villegasSuperconductingReversibleRectifier2003,swiecickiStrongFieldmatchingEffects2012}, further suggesting the dominating role of the vortex in the nonreciprocal transport.

To evaluate the strength of MCA in this system, the ratio of the nonlinear resistance to the linear resistance, $\eta = 2\times R_{2\omega}/R_{\omega}$, under several magnetic fields is plotted in Fig.~\ref{fig:3}f (more data in SM Fig. S6~\cite{SeeSupplementalMaterial}). The magnitude of $\eta$ can reach the limiting value of $\pm 1$, indicating a strikingly large MCA in this system. we can also calculate the MCA coefficient $\gamma = \frac{1}{\mu_0H_zI_{p}}\frac{2R_{2\omega}}{R_\omega}$.
To estimate $\gamma$, we use the values of $R_{\omega}$ and $R_{2\omega}$ under $I_{ac}=\SI{0.8}{\micro\ampere}$ at $\mu_0 H_z$ where $R_{2\omega}$ is at a peak. The $\gamma$ is calculated to be \SI{5.9e8}{\per\tesla\per\ampere} and then $\gamma W$ (W is the sample width) \SI{1.5e3}{\meter\per\tesla\per\ampere}, much larger than previous observations in other two-dimensional noncentrosymmetric superconductors~\cite{wakatsuki2017nonreciprocal,yasuda2019nonreciprocal,itahashi2020nonreciprocal,zhang2020nonreciprocal, ando2020observation,baumgartner2022supercurrent,wakamura2023gate}. According to a theoretical paper \cite{hoshino2018nonreciprocal}, the MCA coefficient in transition-metal dichalcogenides with trigonal symmetry in the superconducting fluctuation regime is enhanced from its normal state value by an order of $(\frac{\epsilon_F}{k_B T_c})^3$. The exceptionally strong MCA in T$_d$-MoTe$_2$ should be closely related to its relatively small $T_c$ compared to other superconducting systems where nonreciprocal transport has been reported. In addition, the lower symmetry of T$_d$-MoTe$_2$ than trigonal symmetry may make the pinning potentials for vortices highly asymmetric, leading to a large MCA~\cite{wakamura2023gate,zhuControllableStepMotors2003}. Finally, the disorders created by argon etching play important role as pinning centers, namely, more defects are helpful to pin the vortex lattices effectively, thereby achieving larger MCA than in pristine samples~\cite{wakamura2023gate}.

\begin{figure*}
\includegraphics{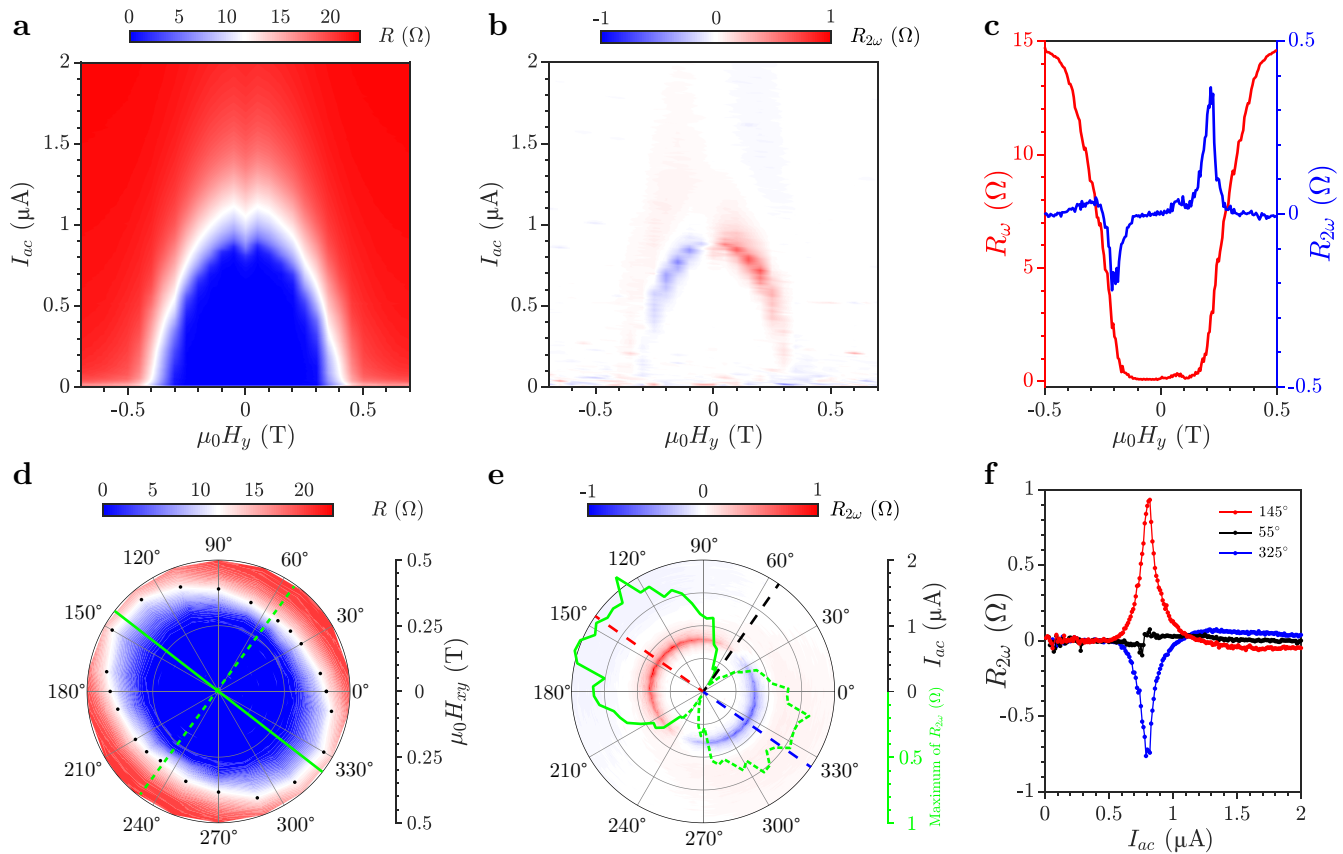}
\caption{\textbf{In-plane magnetochiral anisotropy.} (a) First harmonic resistance $R_\omega$ and (b) second harmonic resistance $R_{2\omega}$ as a function of in-plane perpendicular magnetic field $\mu_0 H_y$ and AC current $I_{ac}$. (c) Field dependences of $R_{\omega}$ and $R_{2\omega}$ under $I_{ac}=\SI{0.65}{\micro\ampere}$. (d) Resistance $R$ as a function of in-plane magnetic field $\mu_0 H_{xy}$ at different azimuthal angle $\varphi$. The black dots indicate the critical magnetic field $\mu_0 H_{c2,\varphi}$ where the reduced resistance $r = 0.5$. The solid green line and dashed line indicate the easy axis and hard axis of the sample. (e) $R_{2\omega}$ as a function of $I_{ac}$ at different $\varphi$ with $\mu_0 H_{xy} = \SI{0.15}{\tesla}$. The solid (dotted) green lines indicate the positive (negative) values. The black, red and blue dashed lines represent the hard axis and the two opposite directions of the easy axis in (d). (f) Linecuts of $R_{2\omega}$ extracted from (e) along the dashed straight lines in their respective colors. All the measurements were performed at \SI{10}{\milli\kelvin}}\label{fig:4}
\end{figure*}

\subsection{In-plane magnetochiral anisotropy}
Now, we turn to the in-plane MCA effect, which is investigated following a similar procedure. Figs.~\ref{fig:4}a and \ref{fig:4}b show the colormaps of the first and second harmonic resistances, $R_\omega$ and $R_{2\omega}$ as a function of the in-plane perpendicular magnetic field $\mu_0 H_y$ and $I_{ac}$, respectively. It is noticeable that $R_{2\omega}$ still appears in the superconducting transition regime. The field dependence of $R_{2\omega}$ under $I_{ac}=\SI{0.65}{\micro\ampere}$ is displayed in Fig.~\ref{fig:4}c, showing an antisymmetric peak/valley structure without the oscillating behavior. The $\eta$ is found to reach a maximal value of 0.5 (SM Fig. S7~\cite{SeeSupplementalMaterial}), indicating the rectification effect is weaker under the in-plane perpendicular magnetic field than the out-of-plane magnetic field. Therefore, the superconducting diode effect is not able to be observed under in-plane magnetic field. 

Next, we rotated the in-plane magnetic field to further characterize the MCA phenomenon. Fig.~\ref{fig:4}d presents the colormap of the magnetoresistance as a function of in-plane field strength and field angle. A significant in-plane anisotropy is observed. The upper critical field $\mu_0H_{c2,\varphi}$, indicated by black dots in the figure, displays a two-fold symmetry with easy axis at $\varphi_a =$ \SI{145}{\degree} and hard axis at $\varphi_b =$\SI{55}{\degree}. The easy axis here refers to the direction where the superconductivity can persist to higher magnetic field \cite{huang2019anisotropic}. The in-plane anisotropy,  $H_{c2,a}/H_{c2,b}$, is determined to be 1.3, and the upper critical field is found to stay above the Pauli-Clogston limit in all in-plane directions, which is calculated to be 318 mT ~\cite{clogston1962upper,chandrasekhar1962note}. Similar observation has been reported in the CVD-grown few-layer T$_d$-MoTe\textsubscript{2} device by another group~\cite{cui2019transport}, which is a result of an asymmetric spin–orbit coupling induced by the breaking of the out-of-plane mirror symmetry and the breaking of the in-plane mirror symmetry along the crystallographic $a$-axis (Fig.~\ref{figDVI}a). In comparison to the results in Ref.~\cite{cui2019transport}, we thus identify the easy axis as the crystallographic $a$-axis. 

Fig.~\ref{fig:4}e shows dependence of $R_{2\omega}$ on the AC current and the field angle $\varphi$, at a constant field strength of \SI{0.15}{\tesla} (see SM Fig. S8 for corresponding $\eta$~\cite{SeeSupplementalMaterial}). And the linecuts of $R_{2\omega}$ along the dash lines in Fig.~\ref{fig:4}e in respective colors are shown in Fig.~\ref{fig:4}f. The green curve in Fig.~\ref{fig:4}e plots the maximal values of $|R_{2\omega}|$ at different field angles $\varphi$, where the solid (dashed) segments indicate positive (negative) values of $R_{2\omega}$. The overall lineshape appears to be symmetric about the easy axis. It is interesting to note that, the maximal values of $R_{2\omega}$ appear to be obtained for magnetic fields parallel to the easy axis, instead of for fields perpendicular to the current flow direction as predicted and reported for Rashba superconductors~\cite{wakatsuki2018nonreciprocal, hoshino2018nonreciprocal,yasuda2019nonreciprocal,itahashi2020nonreciprocal,ando2020observation}. Due to the highly sensitive dependence of superconductivity and MCA on the out-of-plane field, one may argue that the anisotropic $\mu_0H_{c2,\varphi}$ and $R_{2\omega}$ may come from the residual out-of-plane magnetic field if the sample plane was not aligned with the plane where the magnetic fields apply. In this case, the minimum $\mu_0H_{c2}$ along hard axis, corresponding to the largest residual out-of-plane magnetic field, should correspond to the maximum $R_{2\omega}$, which is contradict to our observation of the disappearing $R_{2\omega}$ in Fig.~\ref{fig:4}e. Therefore, the in-plane magnetic field MCA is not caused by the residual out-of-plane magnetic field. To the best of our knowledge, this peculiar behavior with the maximum $R_{2\omega}$ not occurring for fields perpendicular to the current has not been reported before. As mentioned previously, this sample exhibits 2D superconductivity, implying that the coherence length is larger than the film thickness, so the in-plane MCA from vortex effect could be neglected. In the next section, we shall show how the form of the SOC in T$_d$-MoTe$_2$ plays a critical role in determining the angular dependence of the non-reciprocal transport. 

\section{Discussion}
\begin{figure}[t]
    \includegraphics[width=8.cm]{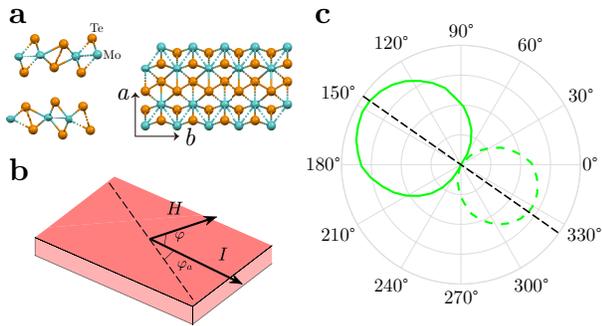}        
    \caption{\textbf{Crystal structure and the calculated nonreciprocal component.} (a) Crystal structure of T$_d$-MoTe$_2$ (left) in side-view and (right) in top-view. The Mo atom and Te atom are represented by balls with different colors. the crystallographic $a$-axis is parallel to the mirror symmetry plane of T$_d$-MoTe$_2$. (b) Illustration of device configuration. The dot-dashed black line indicates the crystallographic $a$-axis. $\varphi_a$ denotes the direction of the a-axis with respect to the current whose flow direction is defined as \SI{0}{\degree}.  $\varphi$ denotes the orientation of the external magnetic field $H$. (c) Angular variation $\varphi$ of the nonreciprocal component, i.e. the DVI in arbitrary unit, with magnetic field orientation, for the crystallographic $a$-axis pointing at $\varphi_a=$\SI{145}{\degree}. Band parameters are set to $(t_1, t_2, \mu, \alpha_1, \alpha_2, \alpha_3)=(2, 4, 1, 0.05, 0.001, -0.01)$.\label{figDVI}
    } 
    \label{fig:FigDVI}
\end{figure}
We now develop a phenomenological theory to explain how an anisotropic SOC may give rise to the peculiar field angle dependence of the nonreciprocal resistance $R_{2\omega}$. The nonreciprocal charge transport in inversion-symmetry breaking systems is rooted in the asymmetric Zeeman shift of the spin-orbit coupled electronic structure~\cite{rikken2001electrical,wakatsuki2018nonreciprocal,ilic2022theory,daido2022intrinsic,yuan2022supercurrent}. This asymmetry causes a disparity between left and right moving electrons, which then leads to inequivalent transport behavior for left and right injected currents. It is thus tempting to describe the nonreciprocal coefficient as proportional to the field-induced differential change in the integral of the group velocity component parallel to the current flow. For later convenience, we shall refer to such a quantity as {\it Differential Velocity Integral} (DVI). In conjunction with our experiments, let's consider a single-crystal device configuration with a DC probe current whose flow direction we set to be $0^\circ$, and with certain main crystallographic axis oriented at angle $\varphi_a$ (see Fig.~\ref{fig:FigDVI}b for illustration). The field angle dependence of the DVI is then given by,
\begin{equation}\label{eqDVI}
    \gamma_{\varphi_a}(\varphi) \propto -\partial_{\mathbf H} \left\{\sum_n \int d\mathbf k v_{n,I}(\mathbf k,\mathbf H) \delta[\xi_{n}(\mathbf k,\mathbf H)] \right\}_{\mathbf H=0} \,.
\end{equation}
Here, $\mathbf H=H(\cos\varphi,\sin\varphi,0)$ denotes the in-plane external field, $\xi_n(\mathbf k,\mathbf H)$ is the dispersion of the $n$-th band and $v_{n,I}(\mathbf k,\mathbf H)$ the group velocity of the same band projected into the current flow direction. Note that the minus sign in the front is due to the negative charge carried by electrons, and that the $\delta$-function reflects the dominant contribution from states around the Fermi level. As an important remark, a serious transport theory would also need to account for the field-induced corrections to other quantities, such as the transport effective mass. However, those corrections are at best the same order as the DVI. For our illustrative purpose, it thus suffices to use the DVI to roughly capture the observed field angle dependence. 

For a simple Rashba SOC model with continuous basal plane rotational symmetry, the DVI is insensitive to $\varphi_a$, but must undergo a sign change when the in-plane magnetic field is (anti)-parallel to the direction of the injected current and reach extremal values when the field is orthogonal. This is indeed verified in the Supplementary Material, and is consistent with the original predictions~\cite{rikken2005magnetoelectric,wakatsuki2018nonreciprocal}. Also presented there is a separate analysis for Dresselhaus SOC, which behaves in the opposite manner. In systems with anisotropic band structure and more complex form of spin-orbit texture, the nonreciprocal coefficient also acquires a dependence on $\varphi_a$, and the DVI develops a more complicated dependence on the field angle $\varphi$. In the following, we turn to the T$_d$-MoTe$_2$ thin film. 

The non-centrosymmetric crystal structure of T$_d$-MoTe$_2$ is characterized by an in-plane mirror symmetry ($\mathcal{M}_y$) while breaking another in-plane and the out-of-plane mirror symmetries (Fig.~\ref{fig:FigDVI}a). This gives rise to an asymmetric SOC \cite{cui2019transport}. In a simplified description, a 2D Hamiltonian appropriate for thin films of this compound, in accordance with the crystallographic and time-reversal symmetries, can be written as,
\begin{equation}\label{eqSOC}
    H_\mathbf k =  \epsilon_\mathbf k + \alpha_1 k_y \sigma_x +\alpha_2 k_x \sigma_y +\alpha_3 k_y \sigma_z \,.
\end{equation} 
Here, $\epsilon_\mathbf k = t_1 k_x^2 + t_2 k_y^2 -\mu$ depicts the spin-independent kinetic energy, and the remaining three terms with Pauli matrices $\sigma_i$ $(i=x,y,z)$ describe the spin-orbit interaction. Assuming $\varphi_a=$\SI{145}{\degree}, Fig.~\ref{fig:FigDVI}c shows the DVI calculated for in-plane fields and under a specific set of band parameters. We see that the lineshape roughly matches with the observed angular variation of $R_{2\omega}$ in Fig.~\ref{fig:4}e, with the two lobes neither perpendicular nor parallel to the DC current. To the best of our knowledge, such a peculiar lineshape has not been reported in other material platforms. To explain this behavior, the form of the SOC is of paramount importance. For one, it is important to steer away from the Rashba limit with $\alpha_1=-\alpha_2$, so that the most pronounced nonreciprocity under in-plane fields does not occur when the field is normal to the current flow direction. For another, a finite $\alpha_3$ is crucial to obtaining unequal (absolute) nonreciprocal coefficient between arbitrary field angles $\varphi$ and $\varphi+180^\circ$, i.e. unequal lobe size as seen in our $R_{2\omega}$ plot (Fig.~\ref{fig:4}e). In Supplemental Material Figs. S9 and S10, the angular variation of the DVI for some other $\varphi_a$ and some other sets of SOC coefficients are plotted, showing strong sensitivity to these parameters~\cite{SeeSupplementalMaterial}.

\section{Conclusion}
In summary, we observed a superconducting diode effect and a large superconducting MCA in noncentrosymmetric type-II Weyl semimetal T$_d$-MoTe\textsubscript{2}. By introducing disorders, the rectification efficiency is greatly enhanced due to the vortex ratchet effect, thereby allowing the superconducting diode effect observed under out-of-plane magnetic field. Contrary to the finding in Rashba superconductors, the strongest MCA effect does not occur when the in-plane magnetic field is perpendicular to the current flow direction. And a phenomenological theory based on asymmetric SOC was developed to explain the result. Inspired by the enhanced nonreciprocal transport through the vortex ratchet effect, it would be interesting to control the disorders better in T$_d$-MoTe\textsubscript{2} by doping, and it is also possible to realize the superconducting diode effect under in-plane field in a thicker sample. Our results reveal the crucial role of the crystallographic symmetry in the nonreciprocal transport and will advance the research for designing the superconducting diode with the best performance. 

During the review process we became aware of related experimental work on gate-tunable MCA in few-layer T$_d$-MoTe$_2$~\cite{wakamura2023gate}, and on superconducting diode effect in MoTe$_2$ Josephson junctions~\cite{chen2023superconducting}.

\section{Method}
\subsection{Crystal growth}
 1T$^\prime$-MoTe\textsubscript{2} bulk crystals were synthesized using the conventional chemical vapor transport process \cite{empante2017chemical,wang2021superconducting}. Briefly, high purity Molybdenum (\SI{99.9}{\percent}) and Tellurium (\SI{99.9}{\percent}) powders were mixed at a stoichiometric ratio of $1:2$ with iodine as transport agent, and then sealed in a quartz tube at high vacuum ($<\num{1e-4}$ torr). The reaction was performed in a two-zone furnace with the hot zone kept at $\SI{780}{\celsius}$ and the cold zone kept at $\SI{700}{\celsius}$ for seven days. The tube was then quenched to room temperature in ice to obtain plate-like crystals.

\subsection{Transport measurement}
The $\text{MoTe}_2$ device was put in a dilution refrigerator with the mixing chamber temperature of about \SI{10}{\milli\kelvin} for a standard four-terminal measurement to perform two individual transport measurement operations. One operation is to measure the differential resistance, by coupling a relatively small AC current $\mathrm{d}I=\SI{40}{\nano\ampere}$ to a DC current $I$ and recording the AC voltage $\mathrm{d}V$ and the DC voltage $V$. The DC current with small AC excitation was generated by a DC source (DC205) and a lock-in amplifier (SR830) coupled by a summing amplifier (SIM918). The AC voltage was measured by a SR830 and the DC voltage by a keysight34461A. The resistance was measured by turning off the DC current output ($I=0$). Another operation is to measure the first and second harmonic signals using an pure AC current generated by a SR830. $I_{ac}$ in the figures is effective value of the AC current. The harmonic signals were recorded by two SR830s. The magnetic field was applied by a vector superconducting magnet.

\begin{acknowledgments}
This work was supported by the Key-Area Research and Development Program of GuangDong Province (Grant No.2018B030327001), Guangdong Provincial Key Laboratory (Grant No.2019B121203002), NSFC (Grant No. 11904155), the Guangdong Science and Technology Department (Grant No. 2022A1515011948),  Shenzhen Science and Technology Program (Grant No. KQTD20200820113010023), National Natural Science Foundation of China (Nos. 11864022, 62264010) and Natural Science Foundation of Jiangxi Province, China (No. 20192ACB21014).
\end{acknowledgments}

%

\end{document}